\documentclass[aps,prd,secnumarabic,showpacs,nobibnotes,showkeys,amsmath,amssymb]{revtex4} 
\begin{document}
\title{\Large\bf  Four Dimensional  Supergravity from String Theory}
\medskip
\author{\large B. B. Deo}
\affiliation{\large Department of Physics, Utkal Universsity, Bhubaneswar-751004, India}
\date{\today}
\begin{abstract}
A derivation of N=1 supergravity action  from string theory is presented. Starting from a 
Nambu-Goto bosonic string, matter field is introduced to obtain a superstring in four 
dimension. The excitation quanta of this string contain graviton and the gravitino. 
Using the principle of equivalence, the action in curved space time are found and the 
sum of them is the Deser-Zumino $N$=1 supergravity action. The energy tensor is Lorentz invariant
due to supersymmmetry.
\end{abstract}
\pacs{04.65.+e, 12.60.J}
\keywords{Supergravity, supersymmetry, superstring}
\maketitle

\section{Introduction}
There are only a few generally accepted ways to extract gravity from String Theory.One of them 
is to follow the method of Feynman\cite{a1} using gravity as a spin-2 field theory coupled to 
its energy momentum tensor. Parallel to this is also the work of Weinberg \cite{a2} who 
constructed a Hamiltonian in interaction picture and obtained the Einstein equation from the 
equations in Heisenberg picture. Another popular method is that of Callan, Fiedan et al \cite{a3} 
who looked for the conditions of consistency for space time to allow propagation of strings. 
Gravitation tensor also emerges in the zero slope limit of the string coupling parameter given 
in reference\cite{a4}.

Here, matter field is first introduced in Nambu-Goto string in 26 dimensions through 11 vector 
Majorana spinors in the bosonic representation of $SO(3,1)$. A superstring is constructed 
which was reported first in reference\cite{a5}. Details of tachyonlessness in mass spectrum,
modular invariance and anomaly cancellation are given in reference\cite{a6}.The particle spectrum
is very rich. 
The construction of the standard model particle spectrum has been reported in reference\cite{a7} 
and the gauge symmetry $SO(6)\otimes SO(5)$ generators along with the descent to three 
generations of standard model, i.e., $Z_3\otimes SU_C(3)\otimes SU_L(2)\otimes U_Y(1)$ have 
also been reported\cite{a8}.

The mass spectrum in the bosonic NS sector\cite{a9} as well as in the fermionic R sector\cite{a10}
contain tachyons. Their self energies cancel and do not show up in the supersymmetric Fock space\cite{a6}.
These tachyons are exceedingly useful in constructing zero mass ground states, which are physical and observable.
In this paper, they are used to construct physical graviton and gravitino in section 3 and 4 respectively.In
sections 5 and 6, action for them in curved space time are derived. In section 2, the quantisation,
Virasoro algebra and physical state conditions are stated. It would be worthwhile to present the salient features 
of the superstring in this section.

Nambu-Goto action with open string coordinates $X^{\mu}(\sigma , \tau)$ in the world sheet $(\sigma , \tau)$ is

\begin{equation}
S=-\frac{1}{2\pi}\int d^2\sigma~\partial^{\alpha}X^{\mu}\partial_{\alpha}X_{\mu},~~~\mu =0,1,2,...,25\label{e1}
\end{equation}
The central charge of this action is 26 and is cancelled by adding the action of the conformal ghost whose
central charge is -26, independent of dimensionality. The theory is made anomaly free. It has been shown by Mandelstam
\cite{a11} that two Majorana fermions in 1 + 1 dimensions are equivalent to one real boson in the finite volume. It
has also been shown that the equivalence is true on a finite interval or a circle like in Veneziano model if the string
is anomaly free. Taking 44 Majorana fermions to form the matter fields and along with the four boson, one can write an anomaly free action
\begin{equation}
S=-\frac{1}{2\pi}\int d^2\sigma~[\partial^{\alpha}X^{\mu}\partial_{\alpha}X_{\mu} -i\sum_{j=1}^{44}
\bar{\phi}^j\rho^{\alpha}\partial_{\alpha}\phi_j]\label{e2}
\end{equation}
with\\
\begin{equation}
\partial_{\alpha}~=~(\partial_{\sigma}, \partial_{\tau}),~~~~\rho^0~=~
\left (
\begin{array}{cc}
0& -i\\
i& 0
\end{array}
\right ),
~~~~\rho^1~=~\left(
\begin{array}{cc}
0& i\\
i& 0
\end{array}
\right )
{\text{and}}~~~  \bar{\phi}~=~\phi^\dag \rho^0\label{e3}
\end{equation}
Unfortunately the Lorentz invariance of the matter part does not hold good. However, there are also Majorana fermions
$\psi^\mu$ in the bosonic representation of SO(3 , 1) so that the action(2) can be recast to read
\begin{equation}
S=-\frac{1}{2\pi}\int d^2\sigma~[\partial^{\alpha}X^{\mu}\partial_{\alpha}X_{\mu} -i\sum_{j=1}^{11}
\bar{\psi}^{\mu,j}\rho^{\alpha}\partial_{\alpha}\psi_{\mu,j}]\label{e4}
\end{equation}
The central charge is 26. However, this is not supersymmetric. The supersymmetric version is
\begin{equation}
S=-\frac{1}{2\pi}\int d^2\sigma~[\partial^{\alpha}X^{\mu}\partial_{\alpha}X_{\mu} -i
\bar{\psi}^{\mu,j}\rho^{\alpha}\partial_{\alpha}\psi_{\mu,j}
+i\bar{\phi}^{\mu,k}\rho^{\alpha}\partial_{\alpha}\phi_{\mu,k} ]  \label{e5}
\end{equation}
Here $j$ runs from 1 to 6 and, the the positive and negative frequency parts are, as normal, 
$\psi^{\mu,j}=\psi^{(+)\mu,j}~+~\psi^{(-)\mu,j}$, whereas the index $k$ runs from 1 to 5 and
$\phi^{\mu,j}=\phi^{(+)\mu,k}~+~\phi^{(-)\mu,k}$. The negative sign of the latter is absorbed in
creation operators which is allowed by the choice of freedom of Majorana fermions.

To write the supersymmetric transformation of the $SO(6) \times SO(5)$ invariant action(5), one has to
introduce arrays $(e^j,e^k)$ which are rows of ten zeros and only one 1 in the $j$th or $k$th place.
$e^je_j$ = 6 and $e^ke_k$ = 5. The upper index refers to a row and the lower to a column.
The action of equation(5) is invariant under supersymmetric transformations
\begin{equation}
\delta X^{\mu} = \bar{\epsilon}(e^j\psi^{\mu}_j - e^k\phi_k^{\mu})\label{e6}
\end{equation}
\begin{equation}
\delta \psi^{\mu,j} =-i~ \epsilon~ e^j\rho^{\alpha} \partial_{\alpha} X^\mu \label{e7}
\end{equation}
and\\
\begin{equation}
\delta \phi^{\mu,k} =-i~ \epsilon~ e^k\rho^{\alpha} \partial_{\alpha} X^\mu \label{e8}
\end{equation}
 Here $\epsilon$ is a constant anticommuting spinor. In this action, there are only four bosonic
coordinates but forty four fermionic modes. There should not be such mismatch. 

The inconsistency should show up by making two successive SUSY transformations. Indeed the two such transformations
lead to spatial 
translation with the coefficient $a^{\alpha}=2i~\bar{\epsilon}^1 \rho^{\alpha}\epsilon_2 ,$ only if
\begin{equation}
\psi_j^{\mu} = e_j\Psi^\mu~~ \text{and}~~~ \phi_k^{\mu} = e_k\Psi^\mu  \label{e9}
\end{equation}
The superpartner of $X^\mu$ is the $\Psi_\mu$ where,
\begin{equation}
\Psi^\mu=e^j\psi_j^\mu-e^k\phi_k^\mu \label{e10}
\end{equation}
The action(5) is reduced to
\begin{equation}
S=-\frac{1}{2\pi}\int d^2\sigma~[~\partial^\alpha X^\mu \partial_\alpha X_\mu -i\bar{ \Psi}^\mu\rho^{\alpha} 
\partial_{\alpha} \Psi_\mu]\label{e11}
\end{equation}
The bosonic and fermionic modes match. The $\Psi^\mu$ is not only the superpartner of $X^\mu$, but also emits the quanta 
of $\psi_j^\mu$ and $\phi_k^\mu$ while in the sites $j$ or $k$ of the array respectively. The theory is tachyonless
in the physical Fock space, modular invariant with a vanishing partition function, ghost free and anomaly free. These
features have been discussed, in detail, in references\cite{a5} and \cite{a6}.
\section{ Quantisation, super virasoro algebra and physical states}
Quantisation is normal as usual.
\begin{equation}
\text{Coordinates:}~~~ X^{\mu}(\sigma ,\tau)=x^\mu + p^\mu \tau + i\sum_{n\ne 0}\frac{1}{n}\alpha_n^{\mu}
e^{-in\tau}~\text{cos}(n\sigma)\label{e12}
\end{equation}
\begin{equation}
\text{~or,}~ \partial_{\pm} X^{\mu} = \frac{1}{2} \sum_{-\infty}^{+\infty} \alpha_n^\mu
e^{-in(\tau \pm \sigma)}\label{e13}
\end{equation}
\begin{equation}
\text{~with}~ [~\alpha_m ^{\mu},\alpha_n ^{\nu}~] =m~ \delta_{m+n}~ \eta^{\mu\nu}\label{e14}
\end{equation}
\begin{equation}
\text{NS fermions:}~~ \psi_{\pm}^{\mu,j}(\sigma,\tau)=\frac{1}{\sqrt 2}~\sum_{r\in Z+1/2}~b_r^{\mu,j}
~~e^{-ir(\tau \pm \sigma)}\label{e15}
\end{equation}
\begin{equation}
 \phi_{\pm}^{\mu,k}(\sigma,\tau)=\frac{1}{\sqrt 2}~\sum_{r\in Z+1/2}~b_r^{\prime \mu,k}
~~e^{-ir(\tau \pm \sigma)}\label{e16}
\end{equation}
This is called the bosonic sector and the quanta obey the anticommutation relations
\begin{equation}
\{b_r^{\mu,j}, b_s^{\nu,j\prime}\}=\delta_{r+s}~\delta_{j,j\prime}~ \eta ^{\mu \nu},~~
\{b_r^{\prime \mu,k}, b_s^{\prime \nu,k^\prime}\}=\delta_{r+s}~\delta_{k,k^\prime}~ \eta ^{\mu \nu}\label{e17}
\end{equation}
\begin{equation}
\text{R fermions:}~~ \psi_{\pm}^{\mu,j}(\sigma,\tau)=\frac{1}{\sqrt 2}~\sum_{m=-\infty}^{+\infty}~d_m^{\mu,j}
~~e^{-im(\tau \pm~ \sigma)}\label{e18}
\end{equation}
\begin{equation}
\phi_{\pm}^{\mu,k}(\sigma,\tau)=\frac{1}{\sqrt 2}~\sum_{m=-\infty}^{+\infty}~d_m^{\prime \mu,k}
~~e^{-im(\tau \pm~ \sigma)}\label{e19}
\end{equation}
\begin{equation}
\text{with}~~~~~~~~~~\{d_m^{\mu ,j}, d_n^{\nu, j^\prime}\}=~\delta_{m+n} \delta_{j,j^\prime}~\eta^{\mu \nu} \label{e20}
\end{equation}
\begin{equation}
\text{and}~~~~~~~~~~\{d_m^{\prime \mu ,k}, d_n^{\prime \nu, k^\prime}\}=
~\delta_{m+n} \delta_{k,k^\prime}~\eta^{\mu \nu} \label{e21}
\end{equation}

To construct admissible physical states, it is essential to give the super Virasoro generators \cite{a12}.
In the notation of reference \cite{a4},
\begin{equation}
L_m=\frac{1}{\pi}\int_{-\pi}^{\pi}d\sigma~e^{im\pi}~T_{++}\nonumber
\end{equation}
\begin{equation}
=\frac{1}{2}\sum_{-\infty}^{+\infty}~:\alpha_{-n}\cdot\alpha_{m+n}:+\frac{1}{2}\sum_{r\in{Z+1/2}}
\left(r + \frac{m}{2} \right):\left( b_{-r}\cdot b_{r+m}-b_{-r}^\prime \cdot b_{r+m}^\prime \right):~~~\text{NS}\label{e22}
\end{equation}
\begin{equation}
=\frac{1}{2}\sum_{-\infty}^{+\infty}~:\alpha_{-n}\cdot\alpha_{m+n}:+\frac{1}{2}\sum_{-\infty}^{+\infty}
\left(n + \frac{m}{2} \right):\left( d_{-n}\cdot d_{m+n}-d_{-n}^\prime \cdot d_{m+n}^\prime \right):~~~\text{R}\label{e23}
\end{equation}
\begin{equation}
G_r=\frac{\sqrt2}{\pi}\int_{-\infty}^{+\infty}d \sigma~e^{i\tau \sigma}~J_{+}=\sum_{-\infty}^{+\infty}
\alpha_{-n} \cdot \left( e^j b_{n+r,j} - e^k b^\prime _{n+r,k} \right)=\sum_{-\infty}^{+\infty}
\alpha_{-n} \cdot B_{n+r}~~~~\text{NS}\label{e24}
\end{equation}
\begin{equation}
F_r=\frac{\sqrt2}{\pi}\int_{-\infty}^{+\infty}d \sigma~e^{i\tau \sigma}~J_{+}=\sum_{-\infty}^{+\infty}
\alpha_{-n} \cdot \left( e^j d_{n+r,j} - e^k d^\prime _{n+r,k} \right)=\sum_{-\infty}^{+\infty}
\alpha_{-n} \cdot D_{n+r}~~~~\text{R}\label{e25}
\end{equation}
\begin{equation}
\text{where}~~~~~~~~~~~~~B^{\mu}_{n+r}=e^j b^{\mu}_{n+r,j}- e^k b^{\prime \mu}_{n+r,k}\label{e26}
\end{equation}
\begin{equation}
\text{and}~~~~~~~~~~~~~~~D^{\mu}_{n+r}=e^j d^{\mu}_{n+r,j}- e^k d^{\prime \mu}_{n+r,k}\label{e27}
\end{equation}

The super Virasoro algebra is
\begin{equation}
[L_m,L_n]=(m-n)L_{m+n}+\frac{c}{12}(m^3-m)~\delta_{m+n}\label{e28}
\end{equation}
\begin{equation}
[L_m,G_r]=\left( \frac{m}{2}-r \right)G_{m+r}~~~~~~~~~~~~~~~~~~~~\text{NS}\label{e29}
\end{equation}
\begin{equation}
\{G_r,G_s\}=2L_{r+s}+\frac{c}{3}(r^2-1/4)~\delta_{r+s}\label{e30}
\end{equation}
\begin{equation}
[L_m,F_n]=\left( \frac{m}{2}-n \right)F_{m+n}~~~~~~~~~~~~~~~~~~~~\text{R}\label{e31}
\end{equation}
\begin{equation}
\{F_m,F_n\}=2L_{m+n}+\frac{c}{3}(m^2-1)~\delta_{m+n}\label{e32}
\end{equation}
The central charge $c$ for the action of equation(5) is 26.

In this superstring, there are four bosons ($c_B$=4), twenty two transverse fermions ($c_T$=11)
$(\psi_j^{1,2},\phi_k^{1,2})$ spanned by two transverse superfermionic modes $\Psi^{1,2}$ and usual
conformal ghosts $c_{FP}$=-26. The light cone part of the action has $c_{l.c.}$=11, which is exactly
equivalent to the action of the superconformal ghosts $(\beta,\gamma)$ as shown in reference\cite{a6}. They
represent the superfermionic modes $\Psi^{0,3}$ representation of the 22 longitudinal ghost modes
$(\psi^{0,3}_j,\phi^{0,3}_k)$. The total central charge is zero and the superstring is anomaly free.

The physical states are defined as
\begin{equation}
\text{NS~:}~~~ (L_0-1)|\Phi>=0, ~~L_m|\Phi>=0,~G_r|\Phi>=0,~~~~r,m > 0 \label{e33}
\end{equation}
\begin{equation}
\text{R~:}~~~ (L_0-1)|\psi>_{\beta}=(F_0^2-1)|\psi>_{\beta}=0,~L_m|\psi>_{\beta}=0,~F_m|\psi>_{\beta}=0,~m>0\label{e34}
\end{equation}
Here, $\beta$ is the spinor component index.

The mass spectra are obtained from $L_0$.
\begin{equation}
\alpha^\prime M^2_{NS}=-1,-\frac{1}{2},0,\frac{1}{2},1,\frac{3}{2}, . . .~~~~~~\text{NS}\label{e35}
\end{equation}
\begin{equation}
\alpha^\prime M^2_{R}=-1,0,1,2, . . .~~~~~~~~~~~~~~~~~~\text{R}\label{e36}
\end{equation}

The G.S.O. projection\cite{a13} operators $[1+(-1)^F]$ on the states eliminates the half integral values.
The tachyonic energy $<0|(L_0-1)^{-1}|0>$ of the bosonic vacuum is cancelled by $-<0|(F_0+1)^{-1}(F_0-1)^{-1}
|0>_R$ of the Ramond sector. The negative sign is due to the fermionic loop. Such tadpole cancellations have
also been noted by Chattarputi et al\cite{a14}. To be more sure, consider the supersymmetric charge\cite{a6}
\begin{equation}
Q=\frac{1}{\pi}\int_0^{\pi} \rho^0 \rho^{\dagger \alpha}\partial_{\alpha}X^\mu \Psi_{\mu}~d\sigma \label{e37}
\end{equation}
and find the supersymmetry result
\begin{equation}
\sum \{Q^\dagger_{\alpha}Q_{\alpha} \}=2H,~~~\sum_{\alpha}|~Q_{\alpha}| ~\phi_0>|^2 = 2< \phi_0|H| \phi_0> \label{e38}
\end{equation}
The ground state is of zero energy. There are no tachyons. The physical mass spectrum in both sectors are integral numbers of 
Regge trajectory, $\alpha^\prime M^2=0,1,2,...$.

However, the existence of the two tachyonic vacuum is important and crucial. These will be used to build ground states of
zero energy in both NS and R sectors. Since
\begin{equation}
[L_0,\alpha^{\mu}_{-1}]=\alpha^{\mu}_{-1}\label{e39}
\end{equation}
the state $A^\mu=\alpha^{\mu}_{-1}|~0>$ has zero energy, i.e., $L_0 A^\mu=0$. Similarly, the product of the two $b$ ( or
$b^\prime $) quanta
\begin{equation}
A^{\mu \nu}_{jj^\prime}=b^\mu_{-1/2,j}b^\nu_{-1/2,j^\prime}|~0> \label{e40}
\end{equation}
has zero energy. The index $j$ runs from 1 to 6. The quanta specified by them belong to $SO(6)$ sector. The $SO(6)$
has a representation which are four component spinors. These will be needed to construct the gravitino from the 
Ramond sector tachyon. Supergravity will need the graviton and a simple choice is to choose $SO(6)$ sector with
indices $j$ running from 1 to 6. This will be built from the NS bosonic tachyon-vacuum by applying two NS creation operators as
in equation(40).
\section{The graviton}
The gravitons are the quanta of the gravitational field. At present there does not exist any complete and self consistent
theory of gravitation. The reasons are nicely given by Weinberg\cite{a15}, which will also be elucidated here. A tensor
of the type $A^{\mu \nu}=\sum_{jj^\prime}A^{\mu \nu}_{j j^\prime}C_{jj^\prime}$ will, in general, contain a spin-0, a spin-1 and
the spin-2. The last is identifiable with the graviton. The first problem is to write a tensor without the dilaton spin-0
or the anti-symmetric tensor spin-1.

$A^{\mu \nu}$ is not only massless but also physical, since $G_{1/2}$ operating on $A^{\mu \nu}$ i.e.,
$G_{1/2}A^{\mu \nu}$ kills one of the $b_j^\dagger$'s of $A^{\mu \nu}$ and the state becomes a half
integral mass tachyonic vacuum. This can be projected out of the physical space by a GSO projection operation.
The product of two $b_j^{\dagger \mu}b_{j^\prime}^{\dagger \nu}$ is antisymmetric in $j$, $j^\prime$ due to
anticommutivity of the $b_j^{\dagger}$s. So the state will exist if $C_{j j^\prime}$ is antisymmetric
and will vanish if $C_{j j^\prime}$ is symmetric.

If the dilaton state is $\Phi$, then
\begin{equation}
g_{\mu \nu}\Phi=g_{\mu \nu}\sum _{ij}C_{ij}b_i^{\dagger \lambda}b_j^{\dagger \lambda}|~0,p> \label{e41}
\end{equation}
Due to anticommuting $b$'s,
\begin{equation}
\sum_{ij}C_{ij}b_i^{\dagger \lambda}b_j^{\dagger \lambda}=-\sum_{ij}C_{ij}b_j^{\dagger \lambda}b_i^{\dagger \lambda}
=-\sum_{ij}C_{ji}b_i^{\dagger \lambda}b_j^{\dagger \lambda}=0, \text{if} ~~C_{ij}=C_{ji} \label{e42}
\end{equation}
Hence, the choice $C_{ij}=C_{ji}$ ensures that there are no dilatons and thus replaces the gauge choice   $h_{\mu \mu}=0$
The anti-symmetric tensorial  zero mass state is
\begin{equation}
A^{\mu \nu}=\sum_{ij}C_{ij}(b_i^{\dagger \mu}b_j^{\dagger \nu}-b_i^{\dagger \nu}b_j^{\dagger \mu})=-A^{\nu \mu} \label{e43}
\end{equation}
Again, if $C_{ij}=C_{ji}$ then $A^{\mu \nu}=-A^{\nu \mu}=0$. The symmetric traceless tensor, which is non zero,
physical and of zero mass is
\begin{equation}
h^{\mu \nu}(p)=\sum_{ij}C_{ij}(b_i^{\dagger \mu}b_j^{\dagger \nu}+b_i^{\dagger \nu}b_j^{\dagger \mu}-2\eta^{\mu \nu}
b_i^\lambda b_j^\lambda)|~0,p> \label{e44}
\end{equation}
Even though $C_{ij}=-C_{ji}$, using GSO projection
\begin{equation}
\left(1+(-1)^F \right)G_{1/2}h_{\mu \nu}(p)=\left((1+(-1)^F \right)[G_{1/2},h_{\mu \nu}(p)]=0\label{e45}
\end{equation}
This can be taken as the graviton. The commutator is
\begin{equation}
[h^{\mu \nu}(p),h^{\dagger \lambda \sigma}(q)]= f^{\mu \nu , \lambda \sigma}|~C~|~^2 \delta^{(4)}(p-q)\label{e46}
\end{equation}
where
\begin{equation}
f^{\mu \nu, \lambda \sigma}=g^{\mu \lambda}g^{\nu \sigma}+g^{\nu \lambda}g^{\mu \sigma}
-g^{\mu \nu}g^{\lambda \sigma}\label{e47}
\end{equation}
and $|C|^2$, besides $C_{ij}$'s, also include normalization factors.

Formally one can go over to quantum field theory by defining the flat space time Fourier transform as
\begin{equation}
h_{\mu \nu}(x)=\frac{1}{2 \pi^3}\int\frac{d^3p}{\sqrt{2p_0}}\left(h_{\mu \nu}(p)e^{ipx}
+h^\dagger_{\mu \nu}(p)e^{-ipx}\right)\label{e48}
\end{equation}
with the commutator
\begin{equation}
[h^{\mu \nu}(x), h^{\dagger \lambda \sigma}(y)]=\frac{1}{2\pi^3}\int \frac {d^3p}{2p_0} \left( e^{ip(x-y)}-e^{-ip(x-y)}
\right)f^{\mu \nu, \lambda \sigma} \label{e49}
\end{equation}
The Feynman propagator is
\begin{equation}
\Delta^{\mu \nu,\lambda \sigma}(x-y)=<0~|~T \left(h^{\mu \nu}(x)h^{\lambda \sigma}(y) \right)~|~0>=
\frac{1}{2\pi^4} \int d^4p~\Delta_F^{\mu \nu, \lambda \sigma}(p)e^{ip(x-y)}\label{e50}
\end{equation}
where
\begin{equation}
\Delta_F^{\mu \nu,\lambda \sigma}(p)=\frac{1}{2}f^{\mu \nu,\lambda \sigma}\frac{1}{p^2-i\epsilon}\label{e51}
\end{equation}
This is the Feynman propagator of the graviton in interaction representation.

Very unfortunately the tensor $h_{\mu \nu}(p)$ is not Lorentz invariant. A true Lorentz  tensor would have helicities 0 and $\pm1$ as well as $\pm2$. Unless the 0 and $\pm1$ are eliminated, they will enter into the calculation of Lorentz invariant amplitudes.
It is essential that the dictation of the generally covariant content of general relativity, i.e., the quantum theory of
gravitation should be Lorentz invariant. There is a method, pioneered by Feynman\cite{a1}, where one starts out with manifestly
Lorentz invariant calculational rules, and then add or modify them to prevent the unphysical particles with helicities 0 or $\pm1$
in physical states. This has been carried out in a successful manner by Mandelstam, Deser and DeWitt\cite{a16}.

A stringy solution is presented here which may be possible only in this new type of four dimensional superstring due to
its supersymmetry content. Examine the action of Lorentz transformation $\Lambda_{\mu \nu}$ on $h_{\mu \nu}$\cite{a15}
\begin{equation}
h_{\mu \nu}~\rightarrow~\Lambda_{\mu}^{\rho} \Lambda_{\nu}^{\sigma}h_{\rho \sigma}+p_\mu\epsilon_\nu+p_\nu\epsilon_\mu \label{e52}
\end{equation}
In string theory $\alpha_0^\mu=p^\mu$. So, writing
\begin{equation}
h_{\mu \nu}(p)=h_{\mu \nu}~|~0,p>~\rightarrow~ \Lambda_\mu^\rho\Lambda_\nu^\sigma  h_{\rho \sigma}~|~0,p>+~O_{\mu \nu}(p)\label{e53}
\end{equation}
where
\begin{equation}
O_{\mu \nu}(p)=(p_\mu \epsilon_\nu + p_\nu\epsilon_\mu)~|~0,p>\label{e54}
\end{equation}
with
\begin{equation}
O_{\mu \nu}=p_{\mu} \epsilon_{\nu}+p_{\nu} \epsilon_{\mu} \label{e55}
\end{equation}
leads to
\begin{equation}
L_0 O_{\mu \nu}(p)= O_{\mu \nu}(p)\label{e56}
\end{equation}
So the extra term generated by Lorentz transformation is tachyonic. This is not permissible in this superstring theory.
Ramond sector states should cancel these additional tachyons due to supersymmetry and since $L_0=F_0^2$,
\begin{equation}
O_{\mu \nu}(p)=L_0O_{\mu \nu}(p)=F_0^2O_{\mu \nu}(p)\label{e57}
\end{equation}
in Ramond sector. So,
\begin{equation}
F_0O_{\mu \nu}^\alpha(p)=\pm O^\alpha_{\mu \nu}(p)\label{e58}
\end{equation}
In general, one can construct spinorial states $|~0>_\alpha$ such that
\begin{equation}
F_0~|~0>_\alpha=~|~0>_\alpha~,~~~ _{\alpha}<0~|~F_0={-} _{\alpha}<0~|~ \text{ and} \sum_\alpha~|~0>_{\alpha \alpha}<0~|~=1 \label{e59}
\end{equation}
\begin{equation}
O_{\mu \nu}(p)=L_0O_{\mu \nu}~|~0,p>=\sum_\alpha F_0~|~0>_{\alpha \alpha}<0~|~F_0O_{\mu \nu}~|~0,p>\nonumber
\end{equation}
\begin{equation}
=-\sum|~0>_{\alpha \alpha}<0~|~O_{\mu \nu}~|~0,p>=-O_{\mu \nu}(p)=0\label{e60}
\end{equation}
This is like the tadpole-like cancellation of the vacuum energies of the two NS and R sectors.All
tachyonic contributions will cancel out due to supersymmetry. Thus $h_{\mu \nu}(p)$ of equation(44)
is Lorentz invariant with spin-2 only. It should not dilute the Lorentz invariant rules in calculation
of quantum transition amplitudes. This solves the most major problem of the quantum theory of gravitation.
\section{Gravitino in flat and curved space time}
In four flat dimensions, the Rarita-Schwinger\cite{a13} equation for spin-$3/2$ objects is 
\begin{equation}
\epsilon^{\mu \nu\lambda \sigma}\gamma_5\gamma_{\lambda}\partial_{\nu}\psi_{\sigma}^{3/2}
=\left( g^{\mu \nu}\gamma^{\sigma}-g^{\mu \sigma}\gamma^{\nu}+g^{\nu \sigma}\gamma^\mu-\gamma^\mu\gamma^\nu\gamma^\sigma
 \right)\partial_\nu \psi_{\sigma}^{3/2}\label{e61}
\end{equation}
As shown by GSO\cite{a13}, in off shell momentum space this is equivalent to the three equations
\begin{equation}
p^\mu \psi_{\mu}^{3/2}=\gamma^\mu \psi_\mu^{3/2}=0 \label{e62}
\end{equation}
\begin{equation}
(\gamma \cdot p)\psi_\mu^{3/2}=0\label{e63}
\end{equation}
Conditions (62) and (63), when satisfied, ensure that there is no admixture of spin-1/2 component. Spin-1/2 component cannot be
ruled out if only equations(62) are satisfied.

Let the Fock ground state above the Ramond tachyon $|~0,p>_\alpha$ be
\begin{equation}
|~\phi_0>=\alpha^{\mu}_{-1}~|~0,p>u_\mu \label{e64}
\end{equation}
or
\begin{equation}
|~\phi_0>=D_{-1}^\mu~|~0,p>u_{1\mu}\label{e65}
\end{equation}
In both the options $L_0~|~\phi_0>=0$, so that the ground state~$|~\phi_0>$ is massless. Physical state constraints
$L_1~|~\phi_0>=0$ and $F_1~|~\phi_0>=0$ imply that
\begin{equation}
p^\mu u_\mu=0,~~\gamma^\mu u_\mu=0 \label{e66}
\end{equation}
Here, $u_\mu$ is a mixture of spin-3/2 and spin-1/2. Since $L_0~|~\phi_0>$ and $L_0=F_0^2$, we have
\begin{equation}
F_0~|~\phi_0>=0 \label{e67}
\end{equation}
It is well known that $D_0^\mu \sim  \gamma^\mu$ as they satisfy the Dirac algebra\cite{a4}. Further,since
$\alpha_0^\mu=p^\mu$, we have
\begin{equation}
p^{\mu} u_{1\mu}=0 ~~~ \text{and}~~\gamma^{\mu} u_{1\mu}=0 \label{e68}
\end{equation}
$u_\mu$ and  $u_{1\mu}$ are arbitrary spinors satisfying (66) and (68) respectively. Let us choose $u_{1\mu} = \gamma^5 u_\mu$.
Obviously
\begin{equation}
\psi_{\mu}^{3/2}=(1+\gamma_5) u_\mu \label{e69}
\end{equation}
satisfies (66) and (68) i.e.,
\begin{equation}
\gamma^{\mu}\psi_{\mu} ^{3/2}=0,~~~ \text{and}~~~ p^\mu \psi_\mu ^{3/2}=0 \label{e70}
\end{equation}
Since $F_0$ is essentially $\gamma \cdot p$, $F_0~|~\phi_0>$ gives us the necessary spin-1/2 admixture elimination condition
\begin{equation}
F_0 \psi_\mu^{3/2}~~\sim ~~ ( \gamma \cdot p) \psi_\mu^{3/2}=0 \label{e71}
\end{equation}
So, $\psi_\mu ^{3/2}$ is the vector spinor of the Rarita-Schwinger equation(57).

The general relativity curved space time equation can be found by invoking the principle of equivalence\cite{a17}. To every 
curved space time point on the manifold, a tangent space is constructed. The vector indices are $a,b,c,...$ in this space.
The vierbein matrices like $e^a_\mu$ take $x-$ space objects to the tangent space and vice versa.
\begin{equation}
e^{a}_{\mu}e^a_\nu=g_{\mu \nu},~~~ e^a_{\mu}e^{b\mu}=\delta^{ab} \label{e72}
\end{equation}
For completely specifying the spin connections, we need $\omega^{ab}_\mu$, in terms of which we introduce the 
covariant derivative $D_\mu$ such that
\begin{equation}
D_{\mu}e^a_\lambda=0 \label{e73}
\end{equation}
With the definition of $\gamma$ matrices
\begin{equation}
\gamma^{a}e_a^\mu=\gamma^\mu\label{e74}
\end{equation}
the covariant derivative of a spinor $\psi$ is
\begin{equation}
D_\mu \psi=\left( \partial_\mu + \omega^{ab}_{\mu}\sigma_{ab} \right)\psi\label{e75}
\end{equation}
Here, $\sigma_{ab}$ is the antisymmetric product of two $\gamma$-matrices. The principle of equivalence is satisfied if
the ordinary flat space derivative $\partial_{\mu}$ is replaced by $D_\mu$. The Rarita-Schwinger equation, in curved 
space time, is
\begin{equation}
\gamma_\nu \gamma^5 D_\sigma \psi_\rho^{3/2} \epsilon^{\mu \nu \sigma \rho}=0 \label{e76}
\end{equation}
The invariant action of the gravitino becomes\cite{a4}
\begin{equation}
S_{\text{gravitino}}=-\frac{i}{2} \int d^4x~e \bar{\psi}_{\mu}^{3/2}\gamma_\nu \gamma^5 D_\sigma \psi_\rho ^{3/2}
\epsilon_{\mu \nu \sigma \rho}\label{e77}
\end{equation}
where $e=\sqrt{g}$.
\section{Graviton and general relativity}
The massless tensor of the graviton field  $h_{\mu \nu}(p)$, given by equation(44), satisfies $L_0h_{\mu \nu}(p)
=-p^2_{\lambda}h_{\mu \nu}(p)=0$. It follows that, in flat space time,
\begin{equation}
\square h_{\mu \nu}(x)=0\label{e78}
\end{equation}
For simplicity and without loss of generality in flat space time, we take plane wave solutions,
\begin{equation}
h_{\mu \nu}(x)= h_{\mu \nu}(p) e^{ipx}\label{e79}
\end{equation}
It is easy to derive
\begin{equation}
h_\lambda ^{\dagger \mu}h_\nu ^\lambda =\delta^\mu _\nu   \label{e80}
\end{equation}
where the coefficients $C_{ij}$ of equation(44) have been adjusted to take care of the factors of $2\pi$, integers leading
to the coefficient of $\delta^{\mu}_{ \nu}$ to be the one in the r.h.s. of equation(80).
In the tangent space,
\begin{equation}
h_{\mu \nu}(x)=e_\mu^a e_\nu^b T_{ab} \label{e81}
\end{equation}
After some calculation~\cite{a4}, one gets
\begin{equation}
[D_\mu~ ,D_\lambda]h_\nu ^\lambda=e_a^\lambda e_{ \nu b}R^{ac}_{\mu \lambda} T^{cb}\label{e82}
\end{equation}
where the Riemannian curvature tensor $R^{ac}_{\mu \lambda}$  is
\begin{equation}
R^{ac}_{\mu \lambda}=\partial_\mu \omega^{ac}_\lambda + \omega^{ab}_\mu \omega^{bc}_\lambda-(\mu 
\leftrightarrow \lambda)\label{e83}
\end{equation}

Inverse calculated from equation(82) is
\begin{equation}
[D_\mu~,D_\lambda]h^\lambda _\nu=R_{\mu \lambda}h_\nu ^\lambda=0\label{e84}
\end{equation}
This is parallel transport equation. Using the normalisation relation (80), it is found that
\begin{equation}
R_{\mu \nu}=h_\nu ^{\dagger \lambda}R_{\mu \sigma}h_\lambda ^\sigma=h_\nu ^{\dagger \lambda}[D_\mu~,
D_\sigma]h_\lambda ^\sigma \label{e85}
\end{equation}
A recipe to go to the curved space time has been succintly put by Misner, Thorne and Wheeler \cite{a17},``The laws of Physics written
in abstract geometric form, differ in no way, whatsoever, between curved space time and flat space time,
this is guaranteed by and, in fact, is a mere rewording of the equivalence principle."  In flat space
 time $D_\mu = \partial_\mu$  and $[\partial_\mu~,\partial_\nu]=0$. Thus r.h.s. of equation(85) vanishes.
 So anywhere in the universe, with curved or flat space time
 \begin{equation}
 R_{\mu \nu}=0\label{e86}
 \end{equation}
 This is the Einstein equation of general relativity deduced from this new construction of the superstring\cite{a5}.
 The action of the graviton is
 \begin{equation}
 S_{\text{graviton}}=-\frac{1}{2\kappa^2} \int d^4x~eR \label{e87}\end{equation}
 
 where
 \begin{equation}
 R=g^{\mu \nu}R_{\mu \nu}\label{e88}
 \end{equation}
 and $\kappa$ is the gravitational constant, whose square is proportional to the Newtonian constant.

\section{Supergravity action and concluding remarks}
 The $N=1$ supergravity action $S$ in this four dimensional string theory is the sum of $S_{\text{gravitino}}$
 and $S_{\text{graviton}}$
 \begin{equation}
 S=-\frac{1}{2} \int d^4x~e  \left(\frac{1}{\kappa^2}R+ \bar{\psi}_\mu ^{3/2}\gamma_\nu \gamma^5 D_\sigma
 \psi_\rho \epsilon^{\mu \nu \sigma \rho} \right)\label{e89}
 \end{equation}
 The local supergravity transformations that leave it invariant are \cite{a4}
 \begin{equation}
 \delta \psi^{3/2}_\mu=\frac{1}{\kappa}D_\mu \epsilon (x)\label{e90}
 \end{equation}
 and
 \begin{equation}
 \delta^m _\mu=- \frac{i}{2} \kappa \bar{\epsilon}(x)\gamma^m \psi^{3/2}_\mu \label{e91}
 \end{equation}
 This was first written down by Weiss and Zumino \cite{a18} in 1976. The application of supergravity to
 quantum gravity has not been pursued vigorously.

 Unfortunately, besides the breakdown of Lorentz invariance  due to the appearance of spin-0 and spin-1
 objects, the theory of quantum gravity contains infinities from integrals over large virtual momenta.
 In quantum electrodynamics, there are only three types of divergences, and they can be absorbed in mass,
 charge and wave function normalisation factors. But in quantum gravity, there are many more infinities.

  Most of these infinities are from from calculation of self energies. If they are tachyonic in
 nature, or even otherwise, they may be cancelled by the Ramond fermion contributions in supergravity. Much work needs to be done in this direction. Perhaps,one has successfully tackled the Lorentz invariance problem of the quantum
 gravity.

 I have been profited by a discussion with Prof. L.Maharana. I  thank Dr. P.K.Jena for his assistance
 and for going through the manuscript.


\begin{thebibliography}{99}
\bibitem{a1} R.P.Feynman, Acta Physica Polonica {\bf 24} (1963)697
\bibitem{a2} S.Weinberg, Phys. Rev.{\bf B138}(1965)988
\bibitem{a3} C.G.Callan, D.Fridan, E.J.Martinec and M.J.Perry, Nucl. Phys. {\bf B262}(1986)593
\bibitem{a4} M.B.Green, J.H.Schwarz and W.Witten, {\it Superstring Theory}, Vol-I(Cambridge University Press, Cambridge,1987), M.Kaku in
{\it Introduction to Superstrings}(Springer Verlag, New York,1988), R.N.Mohapatra, {\it Unification and Supersymmetry}(Springer Verlag, New York,1986)
\bibitem{a5} B.B.Deo, Phys. Lett. { \bf B557}(2003)115
\bibitem{a6} B.B.Deo and L. Maharana, Int. J. Mod. Phys.{A20}(2005)99
\bibitem{a7} B.B.Deo and L.Maharana, Mod. Phys. Lett.{\bf A19}(2004)1939, Gravitation and Cosmology { \bf10}(2004)195
\bibitem{a8} B.B.Deo,{ \it Three Generations of SUSY Standard Model of Nambu-Goto String},hep-th/0501026 v3(2005)
\bibitem{a9} A.Neveu and J.H.Schwarz, Nucl. Phys.{ \bf B31}(1971)86
\bibitem{a10} P.Ramond, Phys. Rev.{ \bf D3}(1971)2415
\bibitem{a11} S.Mandelstam, Phys. Rev.{ \bf D11}(1975)3026
\bibitem{a12} M.A.Virasoro, Phys. Rev. { \bf D1}(1970)2933
\bibitem{a13} F.Gliozzi, J.Scherk and D.Olive, Nucl. Phys.{ \bf B122}(1977)253
\bibitem{a14} A.Chattataputi, F.Englert, L.Houart and A.Taormina, arXiv hep-th/0207238,0212085
\bibitem{a15} S.Weinberg, { \it Gravitation and Cosmology},John Wiley and Sons, New York,1987
\bibitem{a16} B.S.Dewitt, Phys. Rev.{ \bf 162}(1967)1195,1239, erratum Phys. Rev.{ \bf 171}(1968)1834,\\
	      S.Mandelstam, Phys. Rev.{ \bf 175}(1968)1604, S.Deser and B.Zumino, Phys. Lett.
	     {\bf 62B}(1976)335, ibid {\bf 65B}(1976) 369
\bibitem{a17} C.W.Misner, K.S.Thorne and J.A. Wheeler,{ \it Gravitation}, W.H.Freeman and Co., Sanfransisco,1979
\bibitem{a18} J.Wess and B.Zumino,  Nucl. Phys. {\bf B70} (1974)39
\end{thebibliography}
\end{document}